\documentclass[fleqn,twoside]{article}
\usepackage{espcrc2}
\usepackage{epsfig}
\usepackage{graphicx}
\newcommand{\beq}{\begin{equation}}
\newcommand{\eeq}{\end{equation}}
\newcommand{\be}{\begin{eqnarray}}
\newcommand{\ee}{\end{eqnarray}}
\title{Electron- and neutrino-nucleus scattering}

\author{Omar Benhar \\ ~~~ \\
INFN and Department of Physics, Universit\`a ``La Sapienza'' \\
        Piazzale Aldo Moro, 2. I-00185 Roma, Italy}
       
\begin{document}

\begin{abstract}
I review the main features of the nuclear response extracted from
electron scattering data. The emerging picture clearly shows that 
the shell model does not provide a fully quantitative description  
of nuclear dynamics. On the other hand, many body approaches in which
correlation effetcs are explicitely taken into account lead to a satisfactory 
account of electron scattering observables. The possibility of 
exploiting the knowledge acquired from electron scattering to 
reduce the systematic uncertainty of neutrino oscillation 
experiments is outlined.
\vspace{1pc}
\end{abstract}

\maketitle
%%%%%%%%%%%%%%%%%%%%%%%%%%%%%%%%%%%%%%%%%%%%%%%%%%%%%%%%%%%%%%%%%%%%%%%%%%%%%%%%%%%%%
\section{Introduction}
%%%%%%%%%%%%%%%%%%%%%%%%%%%%%%%%%%%%%%%%%%%%%%%%%%%%%%%%%%%%%%%%%%%%%%%%%%%%%%%%%%%%%
Over the past four decades electron scattering has provided a 
wealth of information on nuclear structure and dynamics. 
Form factors and charge distributions have been extracted from 
elastic scattering data, while inelastic measurements have allowed 
for a systematic study of the dynamic response over a broad range 
of momentum and energy transfer. Finally, with the advent of the last 
generation of continuous beam accelerators, a number of exclusive processes 
have been analyzed with unprecedented precision. Recent
theoretical and experimental developments in the field of electron-nucleus 
scattering are reviewed in Ref. \cite{elbavii}. 

In electron scattering experiments the nucleus il mostly seen
as a target. Studying its interactions with the probe, 
whose properties are completely specified, one obtains information on the 
unknown features of its internal structure. In neutrino oscillation 
experiments, on the other hand, nuclear interactions are exploited to detect 
the beam particles, whose kinematics is largely unknown. 

Using the nucleus as a detector obviously requires that its response 
to neutrino interactions be under control at 
a quantitative level. Fulfillment of this prerequisite is in fact critical to
keep the systematic uncertainty associated with the
reconstruction of the neutrino kinematics to an acceptable level. 

This paper is aimed at providing a summary of the picture of the 
nuclear response emerging from the analysis of electron-nucleus scattering data, and
suggesting a possible strategy to exploit the knowledge acquired from electron
scattering in the analysis of the next generation of high precision neutrino
experiments.

In Section 2 I will briefly review the experimental evidence pointing to the 
inadequacy of the mean field pictutre of nuclear dynamics and discuss the role of 
nucleon-nucleon (NN) correlations. Section 3 is devoted to an overview 
of nonrelativistic nuclear many body theory and its 
applications to electron scattering observables. 
In Section 4 I will outline the possible implementation 
of a state of the art theoretical description of the nuclear response 
in the analysis of neutrino oscillation experiments.
Finally, the conclusions are stated in Section 5.
%%%%%%%%%%%%%%%%%%%%%%%%%%%%%%%%%%%%%%%%%%%%%%%%%%%%%%%%%%%%%%%%%%%%%%%%%%%%%%%%%%%%%
\section{Splendor and miseries of the nuclear shell model}
%%%%%%%%%%%%%%%%%%%%%%%%%%%%%%%%%%%%%%%%%%%%%%%%%%%%%%%%%%%%%%%%%%%%%%%%%%%%%%%%%%%%%
The nuclear shell model is based on the assumption that nucleons in a nucleus 
behave as independent particles moving in a mean field. Within this picture 
the many body Schr\"odinger equation reduces to a single particle problem, whose
solution yields the energies and wave functions associated with 
the one-nucleon states. For example, the shell model ground state of oxygen
consists of a core of four nucleons in $S$-states, i.e. carrying orbital angular 
momentum $\ell=0$, and twelve nucleons in the valence $P$-states, 
i.e. with $\ell=1$. Higher energy levels are not occupied.

Electron scattering experiments aimed at assessing the limits of
applicabilty of the nuclear shell model, pioneered by the Frascati 
group in the mid sixties \cite{frascati} 
and systematically pursued in several laboratories over the past four decades 
(reviews of $(e,e^\prime p)$ experiments
can be found in, e.g., Refs. \cite{elbavii,book}), are mainly based on 
measurements of the cross section of the proton knock out process 
\beq
e + A \rightarrow e^\prime + p + (A-1) \ .
\label{eep}
\eeq

The most striking feature emerging from the analysis of $(e,e^\prime p)$ data 
is that, while the spectroscopic lines corresponding to knock out from shell 
model states are clearly seen, the corresponding strengths are consistently and 
sizably lower than expected, regardless
of the nuclear mass number. 

\begin{figure}[hbt]
\begin{center}
\includegraphics[scale=.45]{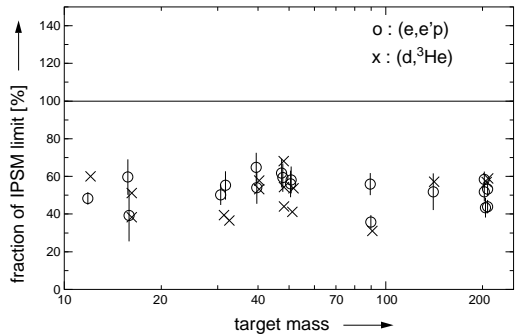}
\end{center}
\vspace*{-.5in}
\caption{\small
Integrated strengths of the valence shell model states, meassured in electron- 
(open circles) and hadron-induced (crosses) proton knock out experiments,
as a function of the target mass number (taken from Ref. \protect\cite{specfact}).
The solid horizontal line represents the shell model prediction.
}
\label{fig:1}
\end{figure}

Fig. \ref{fig:1} shows a recent compilation 
of the strengths of the valence shell model orbits of a number of nuclei, 
ranging from carbon to lead, measured by both electron- and hadron-induced
proton knock out \cite{specfact}. It clearly appears that all the observed 
strengths are largely below the shell model prediction. 
%%%%%%%%%%%%%%%%%%%%%%%%%%%%%%%%%%%%%%%%%%%%%%%%%%%%%%%%%%%%%%%%%%%%%%%%%%%%
%%%%%%%%%%%%%%%%%%%%%%%%%%%%%%%%%%%%%%%%%%%%%%%%%%%%%%%%%%%%%%%%%%%%%%%%%%%%%%

The data displayed in Fig. \ref{fig:1} demonstrate that a significant fraction of 
the target nucleons {\it do not} behave as independent particles, thus providing one of 
the cleanest signatures of correlation effects. Strong NN interactions give 
rise to virtual scattering processes leading to the excitation of the participating 
nucleons to states of energy larger than the Fermi energy, thus depleting
the shell model states within the Fermi sea.

%%%%%%%%%%%%%%%%%%%%%%%%%%%%%%%%%%%%%%%%%%%%%%%%%%%%%%%%%%%%%%%%%%%%%%%%%%%%
\begin{figure}[hbt]
\begin{center}
\includegraphics[width=11pc]{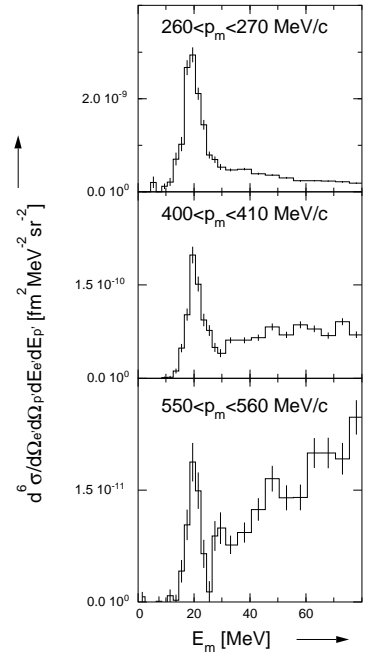}
\end{center}
\vspace*{-.5in}
\caption{\small
Proton removal energy spectra measured in $(e,e^\prime p)$ precesses off
$^4He$ \protect\cite{vanleeuwe98}. The different panels correspond to
different proton momentum ranges.
}
\label{fig:2}
\end{figure}
%%%%%%%%%%%%%%%%%%%%%%%%%%%%%%%%%%%%%%%%%%%%%%%%%%%%%%%%%%%%%%%%%%%%%%%%%%%%%%

To estimate the typical energy scale associated with NN correlations, consider 
a pair of correlated nucleons, carrying momenta ${\bf p}_1$ and ${\bf p}_2$ much 
larger than the Fermi momentum ($\sim 200$ MeV). In the nucleus rest frame, as 
all the remaining $A-2$ particles carry low momenta, 
${\bf p}_1 \approx -{\bf p}_2 = {\bf p}$. Hence, knock out of a nucleon 
of large momentum ${\bf p}$ leaves the residual system with a particle in 
the continuum and requires an energy
\beq
E \approx E_{thr} + \frac{{\bf p}^2}{2m}\ ,
\label{corr:en}
\eeq 
much larger than the Fermi energy ($\sim 30$ MeV).
The above equation, 
where $E_{thr}$ denotes the threshold for two-nucleon removal,
 shows that large separation energy and large nucleon momentum 
are strongly correlated.

Coincidence $(e,e^\prime p)$ experiments have confirmed the validity of the 
simple argument leading to Eq.(\ref{corr:en}). Measurements carried out 
using $^3He$ and $^4He$ targets have clearly shown that, while the knock out of 
a low momentum proton yields an energy spectrum featuring a sharp peak 
corresponding to the transition to a bound state of the residual system, 
the spectra associated with knock out of high momentum nucleons exhibit 
a broad bump, whose maximum is located at an energy roughly given 
by Eq.(\ref{corr:en}) \cite{marchand84,vanleeuwe98}. These features are clearly 
visible in the data shown in Fig. \ref{fig:2} 

A systematic study of proton knock out 
extending to momenta $\sim 700$ MeV and energy $\sim 200$ MeV has been
recently completed at Jefferson Lab using carbon, iron and gold targets. 
Although the data is still being analyzed, the available results appear to confirm 
the presence of an amount of correlated strength consistent with the 
observed depletion of the shell model states \cite{rohe2003,rohe2004}.

%%%%%%%%%%%%%%%%%%%%%%%%%%%%%%%%%%%%%%%%%%%%%%%%%%%%%%%%%%%%%%%%%%%%%%%%%%%%%%%%%%%%%
\section{Many body theory of electron-nucleus scattering}
%%%%%%%%%%%%%%%%%%%%%%%%%%%%%%%%%%%%%%%%%%%%%%%%%%%%%%%%%%%%%%%%%%%%%%%%%%%%%%%%%%%%%

Within the impulse approximation (IA) scheme, which is expected to be applicable at large 
momentum transfer, electron-nucleus scattering is described as
an incoherent sum of elementary scattering processes involving only one 
nucleon, the remaining $A-1$ particle acting as spectators. It follows that, 
neglecting final 
state interactions (FSI) between the struck proton and the residual system, the 
cross section of process (\ref{eep}) can be written in a simple factorized 
form, generally referred to as plane wave impulse approximation (PWIA)
\beq
\frac{d\sigma_{PWIA}}{d\omega d\Omega_{e^\prime} d\Omega_{p^\prime} dT_{p^\prime}}
= |{\bf p}^\prime|(m+T_{p^\prime}) \sigma_{ep} P({\bf p}_m,E_m)\ .
\label{sigmaeep}
\eeq
In the above equation 
 $m$ and ${\bf p}^\prime$ and $T_{p^\prime}$ denote the proton mass, momentum and 
kinetic energy, respectively, while the missing 
momentum ${\bf p }_m$ and missing energy $E_m$ are defined as
${\bf p}_m  = {\bf p}^\prime-{\bf q}$ and $E_m  =  \omega-T_{p^\prime}-T_R$, 
$q$, $\omega$ and $T_R$
being the momentum end energy transfer and the kinetic energy of the recoiling 
spectator system. 

The electron-proton scattering process is described by the elementary
cross section $\sigma_{ep}$, while all the information on nuclear dynamics is
contained in the spectral function $P({\bf p},E)$, defined as 
(see, e.g., ref.\cite{benhar89})
\be
\nonumber
P({\bf p},E) & = & \sum_{n} 
\left| \langle \Psi_n^{(A-1)} | a_{{\bf p}} | \Psi_0^{A} \rangle \right|^2 \\
&  & \ \ \ \ \ \ \ \ \ \ \ \ \ \ \ \times
\delta( E + E_0 - E_n )\ .
\label{sf}
\ee
In the above equation, $| \Psi_0^{A} \rangle$ and $| \Psi_n^{(A-1)} \rangle$ 
describe the target ground
state and the final state of the spectator system, respectively, while $E_0$ and 
$E_n$ are the corresponding energies. The spectral function yields the probability
of removing a nucleon carrying momentum ${\bf p}$ from the target ground state 
leaving the residual system with energy $E$.

Nuclear many body theory (NMBT) provides a fully consistent computational framework
to obtain the spectral function of Eq.(\ref{sf}). Within this approach the nucleus is viewed 
as a collection of pointlike protons and
neutrons, whose dynamics is described by the nonrelativistic hamiltonian
\beq
H = \sum_i \frac{{\bf p}_i^2}{2m} + \sum_{j>i} v_{ij} 
+ \sum_{k>j>i} V_{ijk}\ .
\label{hamiltonian}
\eeq
The two body potential $v_{ij}$ is determined by fitting
deuteron properties and nucleon-nucleon scattering data, while inclusion of the 
three-nucleon interaction is needed to reproduce the binding energy of the three-nucleon systems.

The many body Schr\"odinger equation associated with the hamiltonian
of Eq.(\ref{hamiltonian}) can be solved exactly, using stochastic methods,
for nuclei with mass number up to 10. The resulting energies
of the ground and low-lying excited states are in excellent agreement 
with experimental
data \cite{WP}. Accurate calculations can also be carried out for uniform
nuclear matter \cite{AP}.

\begin{figure}[hbt]
\begin{center}
\includegraphics[scale=.35]{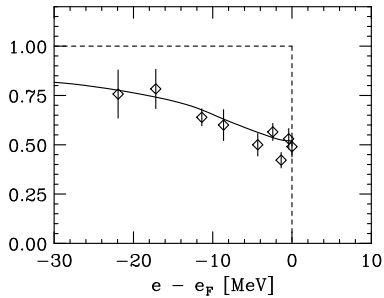}
\end{center}
\vspace*{-.5in}
\caption{\small
Strengths of the single particle states of $^{208}Pb$ measured at NIKHEF, 
plotted as a function of the difference between their energies and the Fermi
energy  \protect\cite{quint88}. The solid line shows the results of the 
theoretical calculation of ref.\protect\cite{benhar90}, based on NMBT,
while the dashed horizontal line corresponds to the shell model prediction.
}
\label{fig:3}
\end{figure}

The spectral functions obtained within NMBT have been extensively used in the 
analysis of a variety of electron scattering observables
(for a review see, e.g., ref.\cite{benhar93_2}). As an example, 
Fig. \ref{fig:3} shows the comparison between the sthengths of the shell model
states of $^{208}Pb$ measured at NIKHEF \cite{quint88} and the theoretical 
results of ref.\cite{benhar90}. It clearly appears that the energy dependence 
of the depletion due to NN correlations is understood at a quantitative level.

%%%%%%%%%%%%%%%%%%%%%%%%%%%%%%%%%%%%%%%%%%%%%%%%%%%%%%%%%%%%%%%%%%%%%%%%%%%%
%%%%%%%%%%%%%%%%%%%%%%%%%%%%%%%%%%%%%%%%%%%%%%%%%%%%%%%%%%%%%%%%%%%%%%%%%%%%%%

The effect of FSI, neglected to obtain Eq.(\ref{sigmaeep}), has long been
recognized to be sizable. Over the past decade
a series of measurements of the nuclear transparency to protons knocked out
in $(e,e^\prime p)$ processes, carried out at MIT Bates \cite{garino92},
SLAC \cite{oneill95} and Jefferson Lab \cite{abbott98,garrow02} have 
consistently shown
deviations of more than 50 \% from the PWIA limit.

Within NMBT FSI effects can be included using the same dynamics 
employed to describe the intial state. In 
the approach developed in ref. \cite{benhar91} the motion
of the knocked out nucleon is treated within the eikonal approximation 
while the spectators are seen as a collection of fixed scattering centers.
Applications to the analysis of inclusive data in the region of
very low energy transfer, which is known to be dominated by FSI
effects, have been quite successful \cite{benhar91,benhar93,benhar94}.

%%%%%%%%%%%%%%%%%%%%%%%%%%%%%%%%%%%%%%%%%%%%%%%%%%%%%%%%%%%%%%%%%%%%%%%%%%%%
\begin{figure}[hbt]
\begin{center}
\includegraphics[scale=.5]{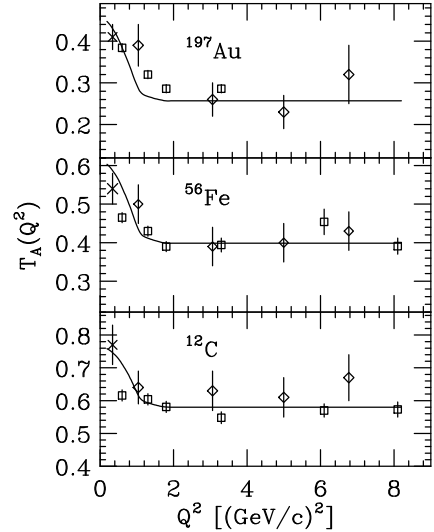}
\end{center}
\vspace*{-.5in}
\caption{\small
$Q^2$-dependence of the transparencies of carbon, iron and gold calculated using 
the local density approximation (LDA) and the approach of 
ref.\protect\cite{benhar91}. The data points are from 
refs.\protect\cite{garino92} (crosses), \cite{oneill95} (diamonds) and
\cite{abbott98,garrow02} (squares).
}
\label{fig:4}
\end{figure}
%%%%%%%%%%%%%%%%%%%%%%%%%%%%%%%%%%%%%%%%%%%%%%%%%%%%%%%%%%%%%%%%%%%%%%%%%%%%%%

Fig. \ref{fig:4} shows the $Q^2$-dependence of the transparency ratio
\beq
T_A(Q^2) = \frac{d\sigma_{expt}}{d\sigma_{PWIA}}\ ,
\label{T}
\eeq 
whose numerator is the observed cross section, whereas the denominator is the 
PWIA cross section og Eq.(\ref{sigmaeep}). From the above definition it
follows that in absence of FSI $T_A(Q^2) \equiv 1$.

The theoretical results have been obtained using the approach 
of ref. \cite{benhar91}, 
whose main ingredients are the measured NN scattering amplitude, corrected to 
take into account medium modifications \cite{pandha92}, and the 
space distribution of the spectator particles, calculated within the local 
density approximation (LDA). 

It has to be pointed out that the
probability of rescattering in the final state does not simply depend upon the 
nuclear density distribution, yielding the probability of finding a spectator
at position ${\bf r}_s$, but upon the {\it joint} probability of finding the struck
particle at position ${\bf r}$ {\it and} a spectator at position ${\bf r}_s$. 
Due to the strongly repulsive nature of NN interactions at short range
this quantity is strongly affected by NN correlations, whose inclusion leads to
a sizable enhancement of the transparency. For example, in lead correlation produce a 
$\sim$ 20 \% effect on $T_A(Q^2)$. 

%%%%%%%%%%%%%%%%%%%%%%%%%%%%%%%%%%%%%%%%%%%%%%%%%%%%%%%%%%%%%%%%%%%%%%%%%%%%%%%%%%%%%
\section{Implementing many body theory in the analysis of neutrino experiments}
%%%%%%%%%%%%%%%%%%%%%%%%%%%%%%%%%%%%%%%%%%%%%%%%%%%%%%%%%%%%%%%%%%%%%%%%%%%%%%%%%%%%%

In the analysis of neutrino experiments nuclear effects are mostly
described using the ultimate independent particle model: the Fermi gas 
(FG) model \cite{moniz69}, according to which the nucleus can be 
approximated by a degenerate gas of protons and neutrons. 

The results of the previous Sections, showing that independent particle
models fail to provide a quantitative account of the nuclear electromagnetic 
response, strongly suggest that correlation effects be large, 
and must therefore be taken into account in the analysys of high precision 
neutrino oscillation experiments.
To see this, consider, for example, the quasielastic charged-current process
\beq
\nu + A \rightarrow \ell + p + (A-1)\ .
\label{cc}
\eeq  
Neutrino kinematics is dictated by energy and momentum conservation, 
requiring
\beq
E_\nu + M_A = E_\ell + E_{p^\prime} + E_{A-1}\ ,
\eeq
$M_A$ being the nuclear mass, and
\beq
{\bf p}_\nu = {\bf p}_\ell + {\bf p}^\prime + {\bf p}_{A-1}\ ,
\eeq
where ${\bf p}^\prime$ denotes the momentum of the outgoing proton, 
\beq
E_{p^\prime} = \sqrt{ {{\bf p}^\prime}^2 + m^2}
\eeq
and
\beq
E_{A-1} = \sqrt{ {\bf p}^2_{A-1} + (M_A - m + E) }\ ,
\eeq
$E$ being the removal energy of the struck nucleon carrying initial 
momentum ${\bf p} \sim -{\bf p}_{A-1}$.

In the FG model all nucleon momenta are lower than the Fermi momentum
$p_F \sim 200$ MeV, while typical removal energies are in the range
$\sim 25 - 35$ MeV \cite{moniz71}. On the other hand, realistic spectral 
functions, on account of NN correlations pushing strength to high 
momentum {\it and} high energy (see Eq.(\ref{corr:en})), 
yield \cite{benhar89,benhar94}
\beq
\langle E \rangle  = \int d^3p\ dE\ P({\bf p},E) \sim 40 - 65\ {\rm MeV}\ , 
\eeq
leading to a neutrino energy sizably different from that obtained using 
the FG model.

The discussion of Section 2 also suggest that the FG model be inadequate
to describe the final state. 
Strong dynamical NN correlations, leading to large
density fluctuations over a length scale of $\sim 1$ fm, have been 
shown to sizably affect 
the probablity of rescattering of the hadrons produced at the weak 
interaction vertex. 

In spite of its success in explaining electron-scattering data, it has
to be realized that exploitation of NMBT in the analysis
of neutrino oscillation experiments, in which the nucleus is seen as a
detector rather than a target, largely depends upon the possibility of 
implementing the theoretical knowledge in Monte Carlo simulations.

%%%%%%%%%%%%%%%%%%%%%%%%%%%%%%%%%%%%%%%%%%%%%%%%%%%%%%%%%%%%%%%%%%%%%%%%%%%%
\begin{figure}[hbt]
\begin{center}
\includegraphics[scale=.42]{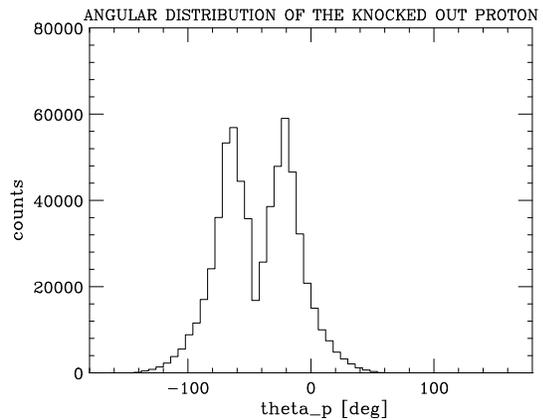}
\end{center}
\vspace*{-.5in}
\caption{\small
Angular distribution of the knocked out proton obtained from a Monte
Carlo simulation of the process $(e,e^\prime p)$ off oxygen at beam energy
700 MeV, energy of the scattered electron 500 MeV and electron scattering
angle 30$^\circ$. The kinetic energy of the ougoing proton is in the
range 160 - 180 MeV.
}
\label{fig:5}
\end{figure}
%%%%%%%%%%%%%%%%%%%%%%%%%%%%%%%%%%%%%%%%%%%%%%%%%%%%%%%%%%%%%%%%%%%%%%%%%%%%%%

Assuming, for the sake of simplicity, that the elementary weak 
interaction vertex in the nuclear medium be the same as in free space,
a realistic simulation of neutrino-nucleus scattering 
requires the energy and momentum probability distribution of
the nucleons, their distribution in space and the medium modified 
hadronic cross section needed for the description of FSI. 

NMBT provides a {\it parameter free} approach that allows one to 
calculate all the above quantities in a fully consistent fashion.

As an example, Fig. \ref{fig:5} shows the proton angular distribution 
resulting from a simulation of the $(e,e^\prime p)$ reaction off oxygen. 
The calculations have been carried out using a spectral function  
obtained within the LDA \cite{benhar04}, ground state configurations 
sampled from the probability distribution associated with the wave
function of ref.\cite{pieper92} and the medium modified NN differential 
cross sections of ref.\cite{pandha92}.
The spectrum of Fig. \ref{fig:5} corresponds to beam energy $E_e = 700$ MeV, 
energy of the scattered electron $E_e^\prime = 500$ MeV and electron scattering 
angle $\theta_e = 30^\circ$. The kinetic energy of the outgoing proton is in 
the range $160 - 180$ MeV.

Thanks to the steady progress of the stochastic techniques to solve the many 
body Scr\"odinger equation, the ingredients needed to carry out realistic
simulations for targets other than oxygen are expected to become available
within the next few years.

%%%%%%%%%%%%%%%%%%%%%%%%%%%%%%%%%%%%%%%%%%%%%%%%%%%%%%%%%%%%%%%%%%%%%%%%%%%%%%%%%%%%%
\section{Conclusions}
%%%%%%%%%%%%%%%%%%%%%%%%%%%%%%%%%%%%%%%%%%%%%%%%%%%%%%%%%%%%%%%%%%%%%%%%%%%%%%%%%%%%%

The results discussed in this paper show that electron 
scattering experiments have exposed the deficiencies of the independent
particle model of nuclear dynamics. On the other hand, many body approaches 
explicitely including dynamical correlation effetcs provide a quantitative
account of a number of electron scattering observables, and appear to 
be a computationally viable option to improve the treatment of nuclear effects
in the analysis of neutrino oscillations experiments. \\ 

\noindent{\bf Acknowledgments}\\

The author is indebted to V.R. Pandharipande, M. Sakuda, R. Seki and
 I. Sick for many illuminating discussions.

%%%%%%%%%%%%%%%%%%%%%%%%%%%%%%%%%%%%%%%%%%%%%%%%%%%%%%%%%%%%%%%%%%%%%%%%%%%%%%%%%%%%%

%%%%%%%%%%%%%%%%%%%%%%%%%%%%%%%%%%%%%%%%%%%%%%%%%%%%%%%%%%%%%%%%%%%%%%%%%%%%%%%%%%%%%
\end{document}